\documentclass{IEEEtran}
\usepackage{framed,multirow}
\usepackage{cite}
\usepackage{amsmath,amssymb,amsfonts}
\usepackage{algorithmic}
\usepackage{graphicx}
\usepackage{textcomp}
\usepackage{multirow}
\usepackage{comment}
\usepackage{tabu}
\usepackage{makecell}
\usepackage[colorlinks=true]{hyperref}
\usepackage{graphicx}
\usepackage[table,xcdraw]{xcolor}
\usepackage[normalem]{ulem}
\useunder{\uline}{\ul}{}

\def\ie{\emph{i.e.}}
\def\eg{\emph{e.g.}}

\def\cf{\emph{c.f.}}

\begin{document}
\title{Structure-Guided MR-to-CT Synthesis with Spatial and Semantic Alignments for Attenuation Correction of Whole-Body PET/MR Imaging}

\author{Jiaxu Zheng$^\dagger$, Zhenrong Shen$^\dagger$, Lichi Zhang*, and Qun Chen
\thanks{$^\dagger$Equally contributed to this work.}
\thanks{*Corresponding author: Lichi Zhang (e-mail: lichizhang@sjtu.edu.cn).}
\thanks{Jiaxu Zheng, Zhenrong Shen, and Lichi Zhang are with the School of Biomedical Engineering, Shanghai Jiao Tong University, Shanghai, China. (e-mail: \{jiaxu.zheng, zhenrongshen, lichizhang\}@sjtu.edu.cn).}
\thanks{Qun Chen is with Shanghai United Imaging Healthcare Co., Ltd., Shanghai, China (e-mail: qun.chen@united-imaging.com).}
}

\maketitle

\begin{abstract}
Deep-learning-based MR-to-CT synthesis can estimate the electron density of tissues, thereby facilitating PET attenuation correction in whole-body PET/MR imaging.
However, whole-body MR-to-CT synthesis faces several challenges including the issue of spatial misalignment and the complexity of intensity mapping, primarily due to the variety of tissues and organs throughout the whole body.
Here we propose a novel whole-body MR-to-CT synthesis framework, which consists of three novel modules to tackle these challenges: 
(1) Structure-Guided Synthesis module leverages structure-guided attention gates to enhance synthetic image quality by diminishing unnecessary contours of soft tissues;
(2) Spatial Alignment module yields precise registration between paired MR and CT images by taking into account the impacts of tissue volumes and respiratory movements, thus providing well-aligned ground-truth CT images during training; 
(3) Semantic Alignment module utilizes contrastive learning to constrain organ-related semantic information, thereby ensuring the semantic authenticity of synthetic CT images.
We conduct extensive experiments to demonstrate that the proposed whole-body MR-to-CT framework can produce visually plausible and semantically realistic CT images, and validate its utility in PET attenuation correction.
\end{abstract}


\section{Introduction}
Simultaneous Positron Emission Tomography and Magnetic Resonance (PET/MR) imaging has been the center of immense focus in research and clinical applications for the past few decades~\cite{quick2014integrated,vandenberghe2015pet}. 
Compared to PET/CT systems, PET/MR imaging substantially reduces the radiation dose, and can integrate the detailed soft tissue information from MR with the radiotracer uptake distribution information from PET for thorough and multi-parametric evaluation.
Therefore, PET/MR systems have been widely applied in various fields of diagnosis tasks such as oncology, neurology, and cardiology, offering deeper insights into disease progression and allowing more precise therapeutic intervention strategies.

The fidelity of PET imaging is inherently susceptible to distortions due to the attenuation of gamma photons when traversing through the human body, which necessitates PET attenuation correction (AC) to ensure the accuracy and reliability of PET imaging~\cite{dekemp1994attenuation}.
CT imaging inherently correlates with the electron density of tissues, making it a natural source for predicting AC maps for PET attenuation correction in PET/CT imaging~\cite{kinahan1998attenuation}.
However, predicting AC maps in PET/MR systems is challenging due to the lack of tissue density information in MR scans.
Therefore, many studies have been investigated for MR-based attenuation correction (MRAC)~\cite{ladefoged2017multi}.
The most straightforward solution is segmentation-based methods, which segment the MR image into different tissue types and assign the corresponding linear attenuation coefficients~\cite{zaidi2003magnetic,hofmann2009towards}.
Another solution is atlas-based techniques, which utilize the registration of MR images and CT templates to produce pseudo-CT images for predicting AC maps~\cite{hofmann2011mri,malone2011attenuation}.
However, both segmentation-based and atlas-based methods face the common challenge of inaccurate tissue density information, particularly the skeletal regions, thus resulting in radiotracer uptake estimation errors in PET image reconstruction~\cite{catana2010toward,keereman2010mri}.
Advancements in deep learning have spurred the research of MR-to-CT synthesis~\cite{nie2017medical,xiang2018deep}.
By learning the intensity mapping between two modalities, MR-to-CT synthesis outperforms conventional MRAC techniques in obtaining tissue density information, providing a vital alternative method for PET attenuation correction in PET/MR imaging~\cite{dong2019synthetic}.

\begin{figure*}[ht]
    \centering
    \includegraphics[width=\textwidth]{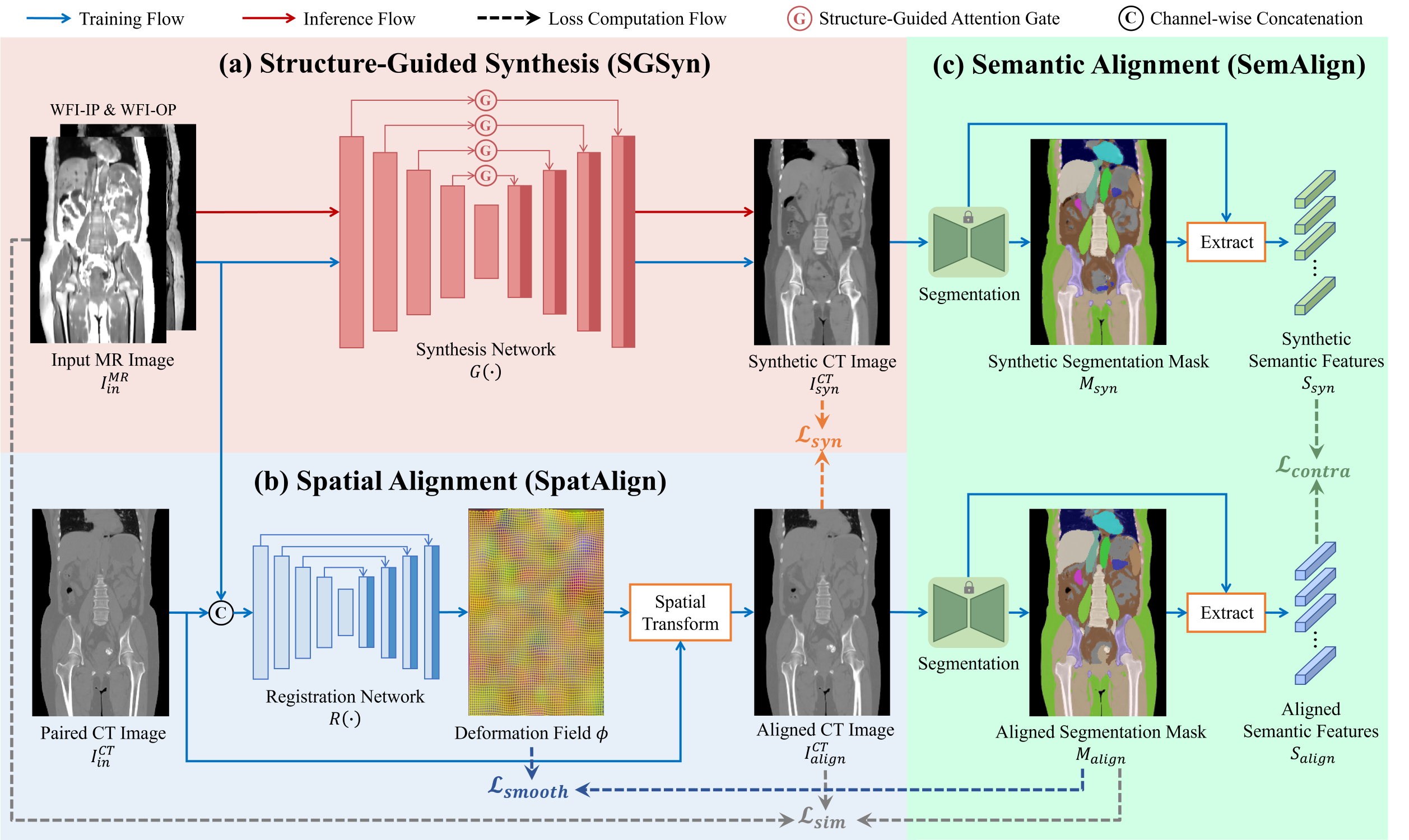}
    \caption{Overview of the proposed whole-body MR-to-CT synthesis framework.
    (a) Structure-Guided Synthesis (SGSyn) translates the input WFI-IP \& WFI-OP MR images $I_{in}^{MR}$ into a synthetic CT image $I_{syn}^{CT}$, which is the main branch of our framework.
    (b) Spatial Alignment (SpatAlign) registers $I_{in}^{MR}$ and the paired CT scan $I_{in}^{CT}$ via a registration network $R(\cdot)$, resulting in the aligned CT image $I_{align}^{CT}$ that provides an explicit similarity constraint on $I_{syn}^{CT}$.
    (c) Semantic Alignment (SemAlign) utilizes contrastive learning to enforce the similarity of identical organs and the dissimilarity of different organs between the feature representation $S_{syn}$ and $S_{align}$, thus ensuring the semantic authenticity of the synthetic results.
    }
    \label{fig:framework}
\end{figure*}

Current studies of MR-to-CT synthesis mainly focus on generating CT images for body sub-regions like head~\cite{ge2019unpaired,yang2020unsupervised} and pelvis~\cite{nie2017medical,hiasa2018cross}, but they fall short in addressing whole-body MR-to-CT synthesis primarily due to two significant challenges:
The \textit{first} challenge lies in the spatial misalignment during the acquisition of paired MR and CT data, which is easily affected by tissue variety and respiratory movements. 
While some research utilizes cross-modality registration techniques in the preprocessing stage, these methods are tailored for specific parts of human body and thus struggle to rectify more intricate whole-body spatial misalignment. 
Other studies propose CycleGAN-based~\cite{zhu2017unpaired} models to directly circumvent the need for registration, but lead to suboptimal synthesis performance due to the lack of explicit image similarity constraints.
The \textit{second} challenge is the complex intensity mapping relationship, an issue stemming from the intensity distribution of soft tissues that show significant variations in MR scans but appear relatively uniform in CT scans.
This complexity is further compounded in whole-body scans due to the variety of tissues and organs.
To address this issue, existing literature utilizes structural features to guide the synthesis process~\cite{hiasa2018cross,yang2020unsupervised} toward high-fidelity results by maintaining structural consistency.
However, they fail to handle more complex intensity mapping as the potential of semantic information inherent in medical images remains largely untapped.

In this paper, we propose a novel whole-body MR-to-CT synthesis framework that converts the in-phase and out-phase MR images of Water-Fat Imaging (WFI-IP \& WFI-OP) into a synthetic CT image for PET attenuation correction.
To achieve high-fidelity MR-to-CT synthesis, the overall framework consists of the following three components as depicted in Fig.~\ref{fig:framework}:
First, we design \textbf{Structure-Guided Synthesis (SGSyn)} module, which incorporates multiple structure-guided attention gates into a 3D synthesis network, aiming at reducing the potential artifacts of anatomical structures in the synthetic CT images.
As the main branch of the proposed framework, SGSyn is implemented in both training and inference stages.
Second, we design \textbf{Spatial Alignment (SpatAlign)} module, which introduces a registration network to handle the spatial misalignment issue between the paired whole-body MR and CT images. 
Here we develop a tissue-aware registration loss to promote the individual alignment of each body sub-region, and a respiration-aware smoothness regularization to make the deformation field conform to respiratory movement.
Third, We design \textbf{Semantic Alignment (SemAlign)} module, which utilizes contrastive learning to draw the semantic features of identical organs in the synthetic and ground-truth CT images closer, while pushing their semantic features of distinct organs further apart.
Note that both SpatAlign and SemAlign are only implemented in the training stage where SpatAlign provides the ground truth image for optimizing the synthesis network and SemAlign complements SGSyn by enhancing the semantic authenticity of the synthetic CT images.

In summary, the main contributions of our work lie in the innovative design of three specialized modules, which collectively address the intricate challenges of whole-body MR-to-CT synthesis for accurate PET attenuation correction in PET/MR imaging, which are listed as follows: 
\begin{itemize}
    \item We design \textbf{SGSyn} which incorporates explicit structural guidance (\ie, edges) to enhance the quality and the fidelity of synthetic CT images; 
    \item We propose \textbf{SpatAlign} which accurately registers CT images to their paired MR scans by accounting for tissue volume variations and respiratory movement influences, thereby addressing the spatial misalignment issue;
    \item We develop \textbf{SemAlign} which leverages the organ-related semantic features to complement \textbf{SGSyn} by ensuring the semantic authenticity of synthetic CT images. 
\end{itemize}

\section{Related Works}
\subsection{MR-based Attenuation Correction}
Attenuation correction is an essential step for PET reconstruction and quantitative analysis, which typically involves the creation of an AC map~\cite{catana2018current}. 
Various MRAC techniques for generating AC maps for PET/MR imaging have been proposed over the past decade, including segmentation-based AC, atlas-based AC, and deep-learning-based AC~\cite{krokos2023review}. 
Specifically, many studies have explored deep learning methods for PET attenuation correction, and have successfully generated patient-specific continuous AC maps for different sub-regions of human body~\cite{gong2018attenuation,arabi2020applications,li2024learning}.

Despite these advancements, deep-learning-based AC methods have not yet become as widely adopted as segment-based AC methods in whole-body PET/MR imaging, mainly due to concerns regarding accuracy and stability~\cite{ahangari2022deep,lindemann2024systematic}.
In addition, segment-based AC methods are inherently deficient in capturing bone information, which often employ atlas-based AC techniques as complementary approaches~\cite{paulus2015whole,arabi2016one}.
Nevertheless, this combined approach yields only discrete tissue attenuation values, and has significant limitations compared to the continuous attenuation correction provided by PET/CT, including artifacts induced by metal implants and patient motion, the impracticality of applying a uniform value to lung tissue, and the discrepancies arising from individual anatomical variations in atlas-based methods~\cite{lindemann2017mr,lillington2020pet,bruckmann2021comparison}.

\subsection{MR-to-CT Synthesis}
Many attempts have been made in the task of MR-to-CT synthesis, yet it remains challenging since paired MR and CT scans typically show spatial misalignment, which also cannot be easily circumvented by CycleGAN~\cite{zhu2017unpaired}. Recent investigations to address this issues can be categorized into the following factors: (1) the incorporation of structural similarity constraints between input and output images~\cite{hiasa2018cross,ge2019unpaired,yang2020unsupervised}, (2) the utilization of attention mechanism by focusing more on structural details than image textures~\cite{kearney2020attention,dovletov2022grad,xiao2024fine}, (3) improvements of the representation for the high-frequency information (\eg, bones) in CT images, where conventional CNNs often find challenging to process effectively~\cite{emami2020attention,shi2021frequency}, and (4) adoptation of auxiliary networks to CycleGAN to enhance the structural information that originates from MR images but might be lost in CT images~\cite{brou2021improving,chen2024icycle}.  
However, these methods are tailored for body sub-regions, which lack specialized image registration modules and thus falling short in addressing the spatial misalignment issue in whole-body MR-to-CT synthesis.

Recently, some studies~\cite{li2023ddmm,ozbey2023unsupervised,pan2024synthetic} also employed diffusion models for MR-to-CT synthesis, which has emerged as a novel paradigm for image generation and demonstrated promising results. 
However, diffusion models are not well-suited for whole-body MR-to-CT synthesis. 
On one hand, these models often struggle with processing large volumes of 3D whole-body scans. 
On the other hand, they are prone to generating anatomical structures that appear realistic but may not be anatomically plausible, posing a significant challenge for applications like PET attenuation correction where the accuracy of tissue densities in synthetic CT images is highly demanded.

\begin{figure*}[ht]
    \centering
    \includegraphics[width=\textwidth]{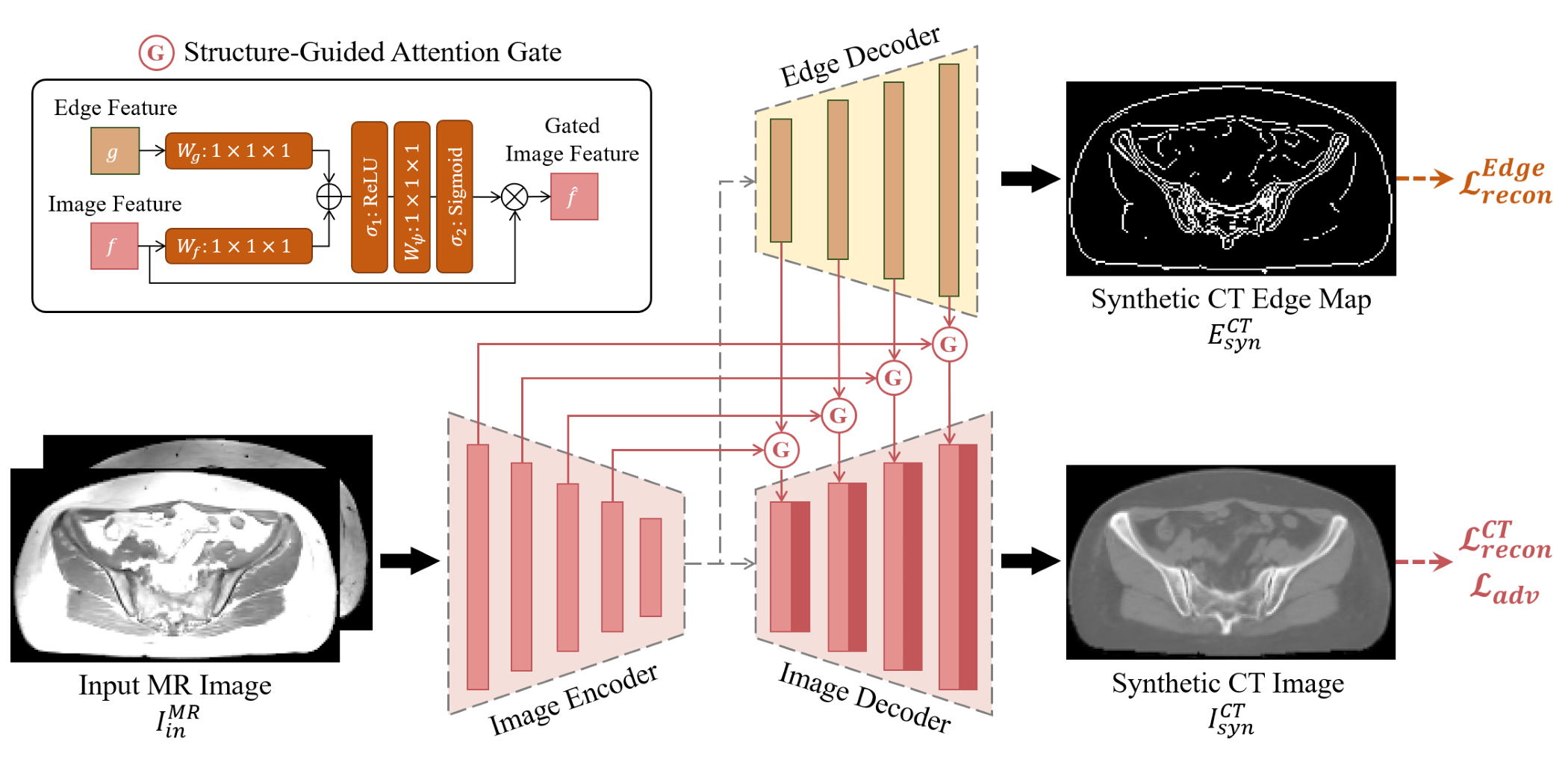}
    \caption{The architecture of the synthesis network from Structure-Guided Synthesis (SGSyn). The network consists of one shared image encoder and two decoders to output a synthetic CT image and its edge map, respectively. It leverages structure-guided attention gates at different scales to facilitate information flow between two decoders for enhancing synthetic CT image quality.}
    \label{fig:gated_unet}
\end{figure*}

\section{Method}
As shown in Fig.~\ref{fig:framework}, here we propose a whole-body MR-to-CT synthesis framework that consists of three modules to synergistically constructing the synthesis network, namely \textbf{Structure-Guided Synthesis (SGSyn)}, \textbf{Spatial Alignment (SpatAlign)}, and \textbf{Semantic Alignment (SemAlign)}. Detailed descriptions of these modules are presented in Section~\ref{sec:synthesis}, Section~\ref{sec:registration}, and Section~\ref{sec:segmentation}, respectively.

\subsection{Structure-Guided Synthesis (SGSyn)}
\label{sec:synthesis}
Soft tissues in a whole-body MR scan typically show significant contrast variations, while a CT scan displays a more uniform intensity distribution across different soft tissues.
Consequently, MR-to-CT synthesis may introduce artifacts to some soft tissues within synthetic CT images.
To address this issue, SGSyn leverages structure-guided attention gates to reduce the incidence of artifacts within the synthetic CT images, thereby enhancing the overall image quality.

As shown in Fig.~\ref{fig:gated_unet}, the synthesis network $G(\cdot)$ employs a modified 3D U-Net~\cite{ronneberger2015u} architecture with one shared image encoder and two decoders to output the synthetic CT image $I_{syn}^{CT}$ and its edge map $E_{syn}^{CT}$, respectively.
The channel-wise concatenation of WFI-IP and WFI-OP MR images is formulated as the input image $I_{in}^{MR}$.
The image encoder downsamples $I_{in}^{MR}$ to extract deep features, while the two decoders with symmetrical architectures gradually upsample the feature map to the original image size.
We also integrate multiple structure-guided attention gates into skip connections between the image encoder and the image decoder, making the network automatically learn optimal feature selection based on structural information from the edge decoder.
For each scale, the structure-guided attention gate can be written as:

\begin{equation}
        \hat{f}=\sigma_{2}\left(W_{\psi}\cdot\left(\sigma_{1}(W_{g}\cdot g + W_{f}\cdot f)\right)\right)\cdot f ,
\end{equation}
where $f$ and $g$ denote feature maps from the image encoder and the edge decoder, respectively.
$W_{f}$, $W_{g}$, and $W_{\psi}$ are channel-wise $1\times1\times1$ convolutions.
$\sigma_{1}$ is the ReLU activation, while $\sigma_{2}$ is the sigmoid activation.
The interaction between $f$ and $g$ produces soft gating weights for each location, which are then assigned across $f$ to obtain the gated image feature $\hat{f}$.
The image decoder aggregates structural guidance from the edge decoder by concatenating $\hat{f}$ at different scales, thus reducing potential artifacts associated with unwanted boundaries.

The training of SGSyn module is implemented with the aid of two loss functions: The voxel-wise $\ell_1$ loss $\mathcal{L}_{recon}^{CT}$ as the reconstruction loss to reduce the intensity difference between the synthetic CT image $I_{syn}^{CT}$ and the ground-truth CT image $I_{gt}^{CT}$, and the $\ell_2$ loss $\mathcal{L}_{recon}^{edge}$ to encourage structure similarity between their edge maps $E_{syn}^{CT}$ and $E_{gt}^{CT}$. The two loss functions are written as:

\begin{equation}
        \mathcal{L}_{recon}^{CT}=\mathbb{E}_{I_{gt}^{CT},I_{syn}^{CT}}\left[{\left\|I_{gt}^{CT}-I_{syn}^{CT}\right\|}_{\ell_1}\right],
\end{equation}
\begin{equation}
        \mathcal{L}_{recon}^{edge}=\mathbb{E}_{E_{gt}^{CT},E_{syn}^{CT}}\left[{\left\|E_{gt}^{CT}-E_{syn}^{CT}\right\|}_{\ell_2}\right],
\end{equation}
where the ground-truth CT image $I_{gt}^{CT}$ is actually $I_{align}^{CT}$ that derives from the paired CT image $I_{in}^{CT}$ via SpatAlign (\cf, Section~\ref{sec:registration}).
The ground-truth edge map $E_{gt}^{CT}$ is obtained from $I_{gt}^{CT}$ using a Canny edge detector.

In addition, to ensure that the intensity characteristics of synthetic CT images match the real data distribution, a 3D fully-convolutional PatchGAN discriminator~\cite{isola2017image} is also employed to distinguish the real CT scans $I_{gt}^{CT}$ from the synthetic ones $I_{syn}^{CT}$ in an adversarial setting, which adopts the standard adversarial loss $\mathcal{L}_{adv}$~\cite{goodfellow2014generative}.
Thus, the total synthesis objective $\mathcal{L}_{syn}$ for this module is formulated as:

\begin{equation}
        \mathcal{L}_{syn}=\mathcal{L}_{recon}^{CT}+\mathcal{L}_{recon}^{edge}+\mathcal{L}_{adv}.
        \label{eq:syn}
\end{equation}

\subsection{Spatial Alignment (SpatAlign)}
\label{sec:registration}
To provide the ground-truth image $I_{gt}^{CT}$ for direct supervision in \eqref{eq:syn}, the proposed SpatAlign produces a spatially well-aligned CT image $I_{align}^{CT}$ by registering the paired CT image $I_{in}^{CT}$ to the input MR image $I_{in}^{MR}$.
Given $I_{in}^{CT}$ as the moving image and $I_{in}^{MR}$ as the fixed image, we design a 3D U-Net registration network $R(\cdot)$ to estimate a deformation field $\phi=R(I_{in}^{MR}, I_{in}^{CT})$, thus enabling the spatial transform to obtain the warped moving image $I_{align}^{CT}=I_{in}^{CT}\circ \phi$.
Referred from VoxelMorph~\cite{balakrishnan2019voxelmorph}, the registration loss $\mathcal{L}_{reg}$ is typically composed of two loss functions as:

\begin{equation}
        \mathcal{L}_{reg}=\mathcal{L}_{sim}(I_{in}^{MR}, I_{align}^{CT})+\lambda\mathcal{L}_{smooth}(\phi),
        \label{eq:reg}
\end{equation}
where the image similarity constraint $\mathcal{L}_{sim}$ maximizes the similarity between $I_{align}^{CT}$ and $I_{in}^{MR}$, and the smoothness regularization $\mathcal{L}_{smooth}$ encourages $\phi$ to be spatially smooth.

Mutual information (MI)~\cite{wells1996multi} is the most widely used image similarity function in multi-modal image registration~\cite{sotiras2013deformable}.
However, directly applying MI as $\mathcal{L}_{sim}$ presents certain complexities due to the non-differentiable nature of its quantization process, \ie, the binning of intensity values.
One feasible solution is to employ Mutual Information Neural Estimation (MINE)~\cite{belghazi2018mutual,snaauw2022mutual} to estimate MI in a continuous quantization manner.
For brevity, we denote the moving image as $\mathbf{m}$, the fixed image as $\mathbf{f}$, and the generated deformation field as $\phi$.
Here MI is equivalent to the Kullback-Leibler (KL) divergence between the joint-probability density $\mathbb{P}=p(\mathbf{f},\mathbf{m} \circ \phi)$ and product of marginals $\mathbb{Q}=p(\mathbf{f})p(\mathbf{m} \circ \phi)$, which can derive a lower-bound MI estimation using the following inequality:

\begin{align}
        D_{KL}&(\mathbb{P} \parallel \mathbb{Q})\ge \nonumber\\
        &\sup_{F_{\theta}\in \mathcal{F}} \mathbb{E}_{\mathbb{P}}\left[F_{\theta}(\mathbf{f},\mathbf{m} \circ \phi)\right]-\mathrm{log}\left(\mathbb{E}_{\mathbb{Q}}\left[e^{F_{\theta}(\mathbf{f},\mathbf{m} \circ \phi)}\right]\right),
        \label{eq:kl}
\end{align}
where $F_{\theta}$ denotes the model parameterized by $\theta$ from the function space $\mathcal{F}$ that satisfies the integrability constraints.
Specifically, the MINE network $F_{\theta}$ is a 3-layer multi-layer perception jointly trained with $R(\cdot)$, and it predicts a single value per voxel pair of $I_{align}^{CT}$ and $I_{in}^{MR}$ to generate the representations of $\mathbb{P}$ and $\mathbb{Q}$.
In this way, MINE shifts the objective of $\mathcal{L}_{sim}$ to maximize the lower bound in \eqref{eq:kl}.
 
As for $\mathcal{L}_{smooth}$, it typically enforces a spatially smooth deformation field $\phi$ by penalizing high-frequency variations to ensure diffeomorphic registration~\cite{balakrishnan2019voxelmorph}:

\begin{equation}
        \mathcal{L}_{smooth}(\phi)=\sum_{\omega\in\Omega}\left\|\triangledown\phi(\omega)\right\|^{2},
        \label{eq:smooth}
\end{equation}
where $\omega$ are points on the image lattice $\Omega$. 
The spatial gradients $\triangledown\phi(\omega)$ are approximated using the differences between neighboring voxels.

Nevertheless, whole-body image registration is challenging as it involves the spatial alignment of multiple tissues and organs.
The structural correspondences are also easily affected by respiratory movements.
Therefore, we specifically modify $\mathcal{L}_{sim}$ and $\mathcal{L}_{smooth}$ by introducing the tissue-aware image similarity constraint and the respiration-aware smoothness regularization, respectively.

\subsubsection{Tissue-Aware Image Similarity Constraint}
Directly applying MINE to calculate $\mathcal{L}_{sim}$ in whole-body image registration can lead to an overemphasis on MI estimation of larger sub-regions due to their higher voxel counts, thereby causing misalignment in smaller sub-regions.
To address this issue, we incorporate a weighted MI term where the weights are inversely proportional to the sub-region volumes. 
Specifically, we formulate a tissue weight map by integrating six distinct sub-regions from the multi-organ segmentation mask $M_{align}$ of the aligned CT image $I_{align}^{CT}$ (\cf, Section~\ref{sec:segmentation}).
Arranged in descending order of volume, these sub-regions include (1) air, as well as organs containing air (\eg, intestines), (2) soft tissues with densities close to water, (3) fat, (4) lungs, (5) pelvis and femur, and (6) ribs and spine.
For a sub-region $i$, we first define its initial weight as the reciprocal of its volume ratio $\frac{\sum_{i=1}^{6} V_{i}}{V_{i}}$. 
Then, we normalize the sum of the weights of all sub-regions, resulting in a tissue weight map.
Finally, we randomly sample $n$ voxel pairs for the MINE network $F_{\theta}$ according to the tissue weight map, thus reaching a tissue-aware image similarity constraint.
Here we empirically set $n=100,000$ in our experiment.

\subsubsection{Respiration-Aware Smoothness Regularization}
The smoothness regularization in \eqref{eq:smooth} is applicable to continuous motions such as the compression of organs when the human body lies on the scanning bed.
However, the respiratory movement during whole-body imaging usually causes slippage between the liver and thoracic cavity, which does not conform to the continuity assumed in the deformation field.
Therefore, we particularly derive the boundaries of thoracic cavity from the segmentation mask $M_{align}$ of the aligned CT image $I_{align}^{CT}$, and exclude the points within this sliding region from calculating $\mathcal{L}_{smooth}$ in \eqref{eq:smooth}:

\begin{equation}
        \mathcal{L}_{smooth}(\phi)=\sum_{\hat{\omega}\in\hat{\Omega}}\left\|\triangledown\phi(\hat{\omega})\right\|^{2},
\end{equation}
where $\hat{\omega}$ are points on the image lattice $\hat{\Omega}$ that eliminates the boundary of thoracic cavity, thereby achieving a smoothness regularization aware of respiratory movement.

\subsection{Semantic Alignment (SemAlign)}
\label{sec:segmentation}
While the synergy between SGSyn and SpatAlign offers an explicit structural similarity constraint on training the synthesis network $G(\cdot)$, the SemAlign module is designed to encourage $G(\cdot)$ to grasp organ-related semantic information by contrastive learning, which can further enhance the semantic authenticity of synthetic CT images.

To obtain semantic representations of various organs, we first employ MOOSE~\cite{sundar2022fully}, a pre-trained multi-organ segmentation model for CT volumes based on nnU-Net~\cite{isensee2021nnu}, to produce a multi-organ segmentation mask $M_{syn}$ from the synthetic CT image $I_{syn}^{CT}$.
The mask $M_{syn}$ is then applied to the feature map from the penultimate layer of MOOSE, thereby extracting semantic features $S_{syn}$ from different organs.
Specifically, let $S_{syn}=\{v_1,v_2,...,v_n\}$, where $v\in\mathbb{R}^{C}$ is a 1-D semantic feature vector extracted from the masked feature map in a voxel-wise manner.
Given a pair of feature vectors $(v_i, v_j)$, they are denoted as a positive pair if extracted from the same organ, otherwise they are considered as a negative pair.
We extract the semantic features $S_{align}=\{v_1,v_2,...,v_m\}$ from the ground-truth image $I_{align}^{CT}$ in the same way.
To push the feature representations of the same organs to be close while distinct organs to be far apart, we implement InfoNCE loss~\cite{he2020momentum} as the contrastive loss $\mathcal{L}_{contra}$:

\begin{equation}
        \mathcal{L}_{contra}=-\frac{1}{\left|\mathcal{P}\right|} \sum_{(v_i, v_j)\in\mathcal{P}}\mathrm{log}\left(\frac{\mathrm{exp}\left(\mathrm{sim}(v_i,v_j)\right)}{\sum_{k=1}^{N}\mathrm{exp}\left(\mathrm{sim}(v_i,v_k)\right)} \right),
        \label{eq:contra}
\end{equation}
where $\mathcal{P}$ is the set of all positive pairs from the union of semantic features $S=S_{syn}+S_{align}=\{v_1,v_2,...,v_N\}$, and $\mathrm{sim}(\cdot)$ represents cosine similarity.
In this way, $\mathcal{L}_{contra}$ not only establishes semantic separability among different organs within the synthetic CT image $I_{syn}^{CT}$, but also identifies the relations of semantic information between the synthetic organs and those in the real image $I_{align}^{CT}$.

\begin{figure*}[ht]
    \centering
    \includegraphics[width=\textwidth]{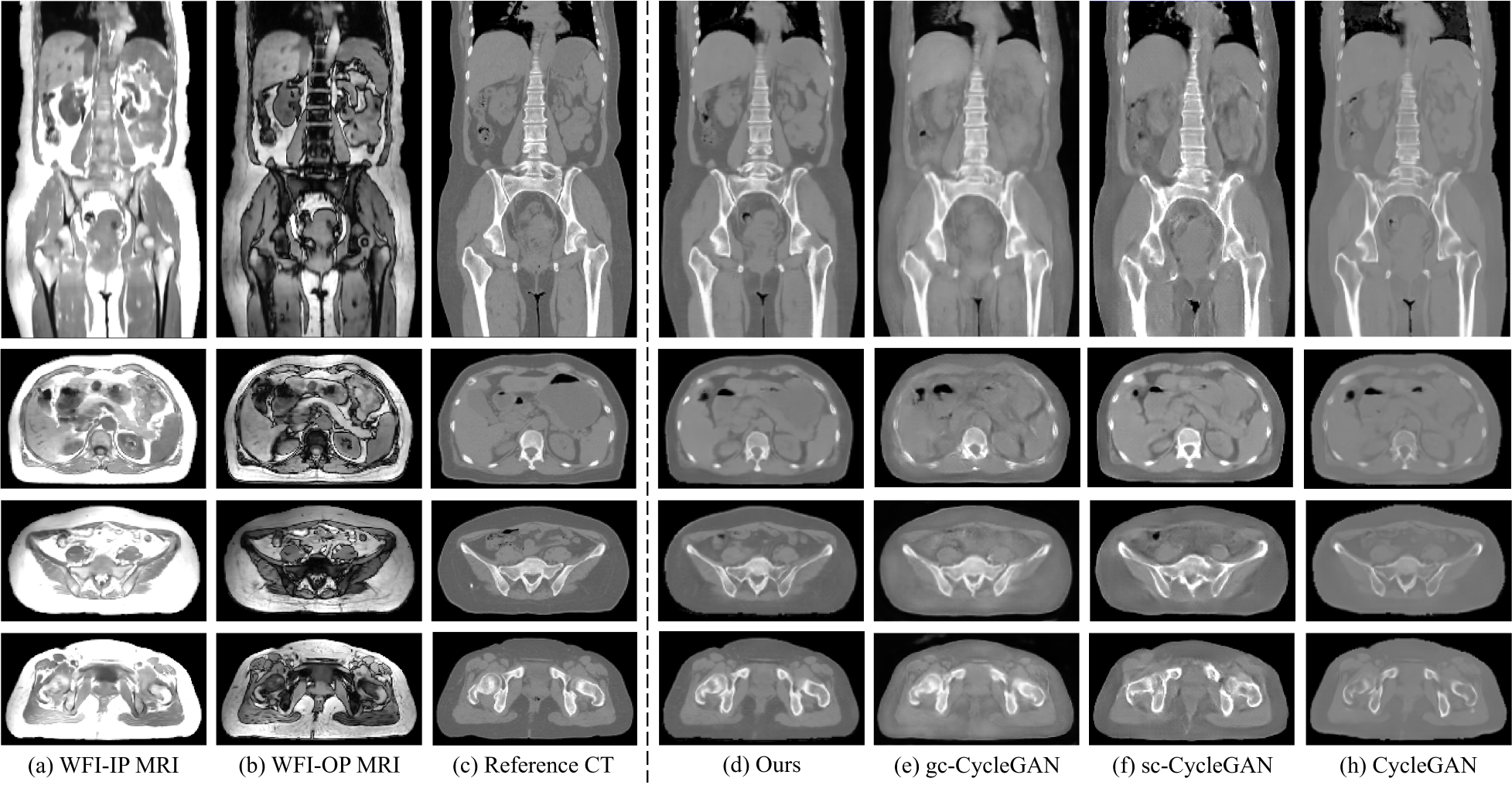}
    \caption{Qualitative comparison between the proposed method and other MR-to-CT synthesis models. The columns from left to right include (a) the input WFI-IP MR image, (b) the input WFI-OP MR image, (c) the reference CT image (not well-aligned with MR images), and the synthetic CT images using (d) our method, (e) gc-CycleGAN~\cite{hiasa2018cross}, (f) sc-CycleGAN~\cite{yang2020unsupervised}, and (h) CycleGAN~\cite{zhu2017unpaired}, respectively.}
    \label{fig:compare}
\end{figure*}

\begin{table*}[htbp]
\centering
\caption{Quantitative comparison between the proposed framework and other methods in whole-body MR-to-CT synthesis. Bold and underlined fonts are the best and second-best results, respectively (↓: Lower is better; ↑: Higher is better).}
\label{tab:compare}
\resizebox{\textwidth}{!}{%
\begin{tabular}{c|ccccc|ccccc}
\hline
\multirow{2}{*}{Method} & \multicolumn{5}{c|}{PSNR↑}                                                       & \multicolumn{5}{c}{SSIM↑}                                                         \\ \cline{2-11} 
                        & Whole Body     & Spine         & Liver         & Ribs           & Femur          & Whole Body     & Spine          & Liver         & Ribs           & Femur          \\ \hline
CycleGAN~\cite{zhu2017unpaired}                & 18.81          & 13.89         & 21.27         & 7.38           & 14.56          & 0.599          & {\ul 0.635}    & 0.609         & 0.481          & 0.511          \\
gc-CycleGAN\cite{hiasa2018cross}             & {\ul 19.16}    & {\ul 14.70}   & {\ul 24.98}   & 8.18           & {\ul 15.58}    & 0.617          & 0.564          & 0.609         & 0.455          & {\ul 0.672}    \\
sc-CycleGAN~\cite{yang2020unsupervised}             & 19.10          & 13.94         & 22.84         & {\ul 8.97}     & 15.45          & {\ul 0.645}    & 0.619          & {\ul 0.659}   & {\ul 0.497}    & 0.654          \\
Ours                    & \textbf{21.71} & \textbf{15.9} & \textbf{22.9} & \textbf{13.25} & \textbf{17.03} & \textbf{0.772} & \textbf{0.714} & \textbf{0.68} & \textbf{0.642} & \textbf{0.731} \\ \hline
\end{tabular}%
}
\end{table*}

\section{Experiments}
\subsection{Dataset and Experimental Setup}
\subsubsection{Dataset}
In this study, our experiments were conducted on a multi-center dataset involving 350 subjects from collaborative hospitals with IRB approvals.
Each subject underwent whole-body PET/MR scanning by UIH uPMR790 scanner to obtain WFI-IP \& WFI-OP MR volumes. 
They also underwent whole-body PET/CT scanning by UIH uMI780 scanner to get CT volumes.
In original data, the MR volumes were $549\times384\times472$ with the spacing of 0.97 mm $\times$ 0.97 mm $\times$ 2.40 mm while the CT volumes were $512\times512\times900$ with the spacing of 0.98 mm $\times$ 0.98 mm $\times$ 1.00 mm.

For pre-processing, the regions above the shoulders have been cropped out entirely to avoid the negative impact of varying arm positions during scanning.
Concurrently, the patient beds that were originally visible in CT scans have also been erased.
Next, all CT and MR volumes were resampled to $128\times128\times128$ with the spacing of 2 mm $\times$ 2 mm $\times$ 2 mm, and were normalized to the intensity range of $[-1, 1]$.
N4 Bias Field Correction~\cite{tustison2010n4itk} was then applied to the MR images to correct their intensity inhomogeneities.
For the initial spatial alignment between paired MR and CT images, affine spatial normalization was performed for linear registration, and then B-spline registration using the Elastix toolbox~\cite{klein2009elastix} was applied to refine the registration results. 
We randomly selected 300 subjects for training while the remaining 50 were for testing.

\subsubsection{Implementation Details}
All experiments in this study were conducted using four NVIDIA GeForce A40 GPUs with PyTorch~\cite{paszke2019pytorch}. 
Specifically, we used the learning rate of $4\times10^{-4}$, batch size of 4, and Adam optimizer~\cite{kingma2015adam} to solely train the registration network $R(\cdot)$ in SpatAlign for 200 epochs.
Then, we froze $R(\cdot)$ to train the synthesis network $G(\cdot)$ using the proposed framework, with the learning rate of $2\times10^{-4}$, batch size of 4, and Adam optimizer for 200 epochs.
Data augmentation techniques ($p=0.2$) were implemented when training $R(\cdot)$ and $G(\cdot)$, including Gaussian filtering, affine transformation, spline warping, and 3D rotation.
The evaluation of Peak Signal-to-Noise Ratio (PSNR) and Structural Similarity (SSIM) were conducted to measure the quality of synthetic CT images, while the standard uptake value (SUV) difference was used to evaluate their PET attenuation correction performance. 

\begin{figure*}[ht]
    \centering
    \includegraphics[width=\textwidth]{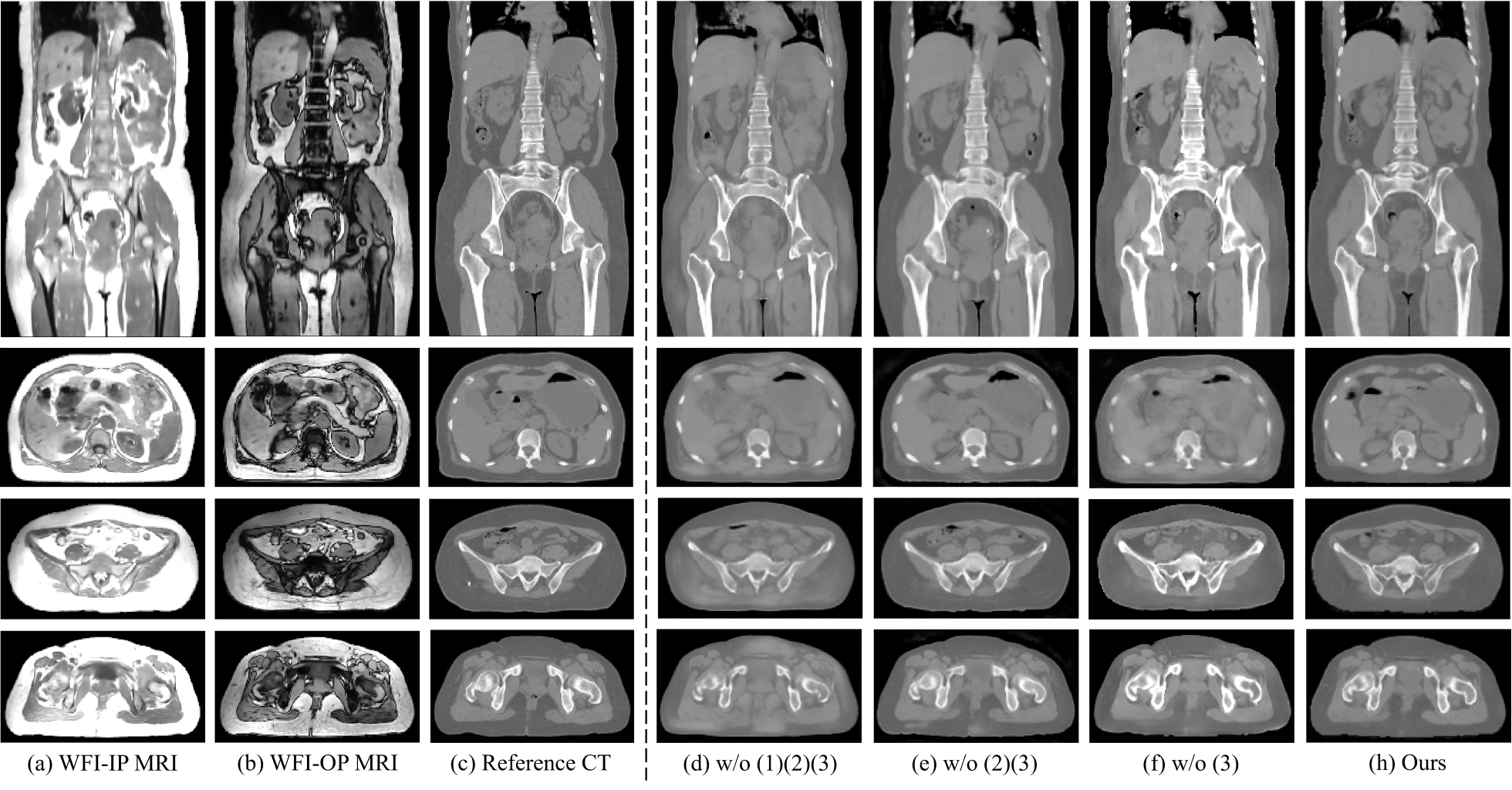}
    \caption{Qualitative results from ablation study of key designs in three modules: (1) the gated U-Net design in SGSyn; (2) the tissue-aware image similarity constraint and the respiration-aware smoothness regularization in SpatAlign; (3) the contrastive loss in SemAlign.}
    \label{fig:ablation}
\end{figure*}

\begin{table*}[htbp]
\centering
\caption{Quantitative ablation study of key designs in each module: (1) the gated U-Net design in Cross-modality Synthesis Module; (2) the tissue-aware image similarity constraint $\mathcal{L}_{sim}$ and the respiration-aware smoothness regularization $\mathcal{L}_{smooth}$ in Spatial Alignment Module; (3) the contrastive loss $\mathcal{L}_{contra}$ in Semantic Alignment Module (↑: higher is better).}
\label{tab:ablation}
\resizebox{\textwidth}{!}{%
\begin{tabular}{ccc|ccccc|ccccc}
\hline
\multicolumn{3}{c|}{Configuration}                & \multicolumn{5}{c|}{PSNR↑}                                                       & \multicolumn{5}{c}{SSIM↑}                                                          \\ \hline
Gated U-Net & $\mathcal{L}_{sim}$ \& $\mathcal{L}_{smooth}$ & $\mathcal{L}_{contra}$ & Whole Body     & Spine         & Liver         & Ribs           & Femur          & Whole Body     & Spine          & Liver          & Ribs           & Femur          \\ \hline
×           & ×                & ×         & 20.96          & 14.05         & 21.90         & 10.27          & 14.16          & 0.730          & 0.581          & 0.582          & 0.562          & 0.634          \\
\checkmark           & ×                & ×         & 21.17          & 13.22         & 22.21         & 10.01          & 13.30          & 0.748          & 0.605          & 0.626          & 0.605          & 0.662          \\
\checkmark           & \checkmark                & ×         & 21.23          & 15.68         & 22.70         & 10.76          & 16.62          & 0.751          & 0.642          & 0.667          & 0.581          & 0.681          \\
\checkmark           & \checkmark                & \checkmark         & \textbf{21.71} & \textbf{15.90} & \textbf{22.90} & \textbf{13.35} & \textbf{17.03} & \textbf{0.772} & \textbf{0.714} & \textbf{0.680} & \textbf{0.642} & \textbf{0.731} \\ \hline
\end{tabular}%
}
\end{table*}

\subsection{Evaluation of MR-to-CT Synthesis}
\subsubsection{Comparison with SOTA Methods}
We first compare the proposed framework with three SOTA methods in whole-body MR-to-CT synthesis task, including (1) CycleGAN~\cite{zhu2017unpaired}, (2) CycleGAN with gradient-consistency loss (gc-CycleGAN)~\cite{hiasa2018cross}, and (3) the structure-constrained CycleGAN (sc-CycleGAN)~\cite{yang2020unsupervised}.
As shown in Fig.~\ref{fig:compare}, the proposed method can predict visually realistic and semantically authentic CT images, with clear structures and plausible intensities for different organs and tissues.
In contrast, CycleGAN only generates CT images that exhibit dim intensities in bones, unaligned organ shapes, and fuzzy tissue boundaries.
While gc-CycleGAN and sc-CycleGAN show improvements in organ spatial alignment due to their structure consistency constraints during training, their synthetic results still suffer from artifacts (\eg, bladder, soft tissues near femur). 
The underlying reason is that they only enforce the spatial structure alignment between the input and output images while neglecting their semantic correspondence.
This leads to the synthesis of unnecessary contours to achieve structural correspondence, thus resulting in unwanted artifacts.
On the contrary, our method not only ensures structural alignment but also guarantees semantic correspondence, thereby avoiding the presence of artifacts.

We also report on their quantitative comparison in Table~\ref{tab:compare} for the whole body and separate organs such as spine, liver, ribs, and femurs.
Our method achieves the highest scores in all metrics across different sub-regions.
This comparison results demonstrate that the direct supervision provided by SpatAlign can effectively promote the fidelity of intensity values and image structures, as opposed to other CycleGAN-based methods that solely rely on adversarial learning to impose image similarity constraints.
The most evident discrepancy can be observed in the rib regions due to their faint appearance in MR images, which poses a great challenge for CycleGAN-based methods.
Conversely, they only show higher performance in the liver region, which is obvious in MR images but still falls behind our method.
In summary, our proposed framework outperforms all other approaches in both qualitative and quantitative comparisons, demonstrating its superiority in whole-body MR-to-CT synthesis.

\begin{table*}[htbp]
\centering
\caption{Quantitative comparison between the proposed framework and other methods in PET attenuation correction. Bold fonts are the best results in terms of the mean SUV differences. (↓: lower absolute value is better)}
\label{tab:pet_recon}
\resizebox{\textwidth}{!}{%
\begin{tabular}{c|ccccc}
\hline
\multirow{2}{*}{Method} & \multicolumn{5}{c}{SUV Difference↓}                                                                                                       \\ \cline{2-6} 
                        & Spine                     & Liver                     & Thigh                    & Pelvis                    & Femur                     \\ \hline
Segmentation-based~\cite{keereman2010mri}      & -0.1314 ± 0.0608          & -0.0142 ± 0.0120          & -0.0193 ± 0.0238         & -0.0721 ± 0.0834          & -0.2626 ± 0.0563          \\
sc-CycleGAN~\cite{yang2020unsupervised}             & -0.0541 ± 0.0801          & -0.0071 ± 0.0059          & \textbf{0.0019 ± 0.0092} & \textbf{-0.0421 ± 0.0805} & -0.0802 ± 0.0673          \\
Ours                    & \textbf{-0.0509 ± 0.0326} & \textbf{-0.0018 ± 0.0042} & 0.0021 ± 0.0041          & -0.0493 ± 0.0304          & \textbf{-0.0478 ± 0.0383} \\ \hline
\end{tabular}%
}
\end{table*}

\begin{figure}[t]
    \centering
    \includegraphics[width=\linewidth]{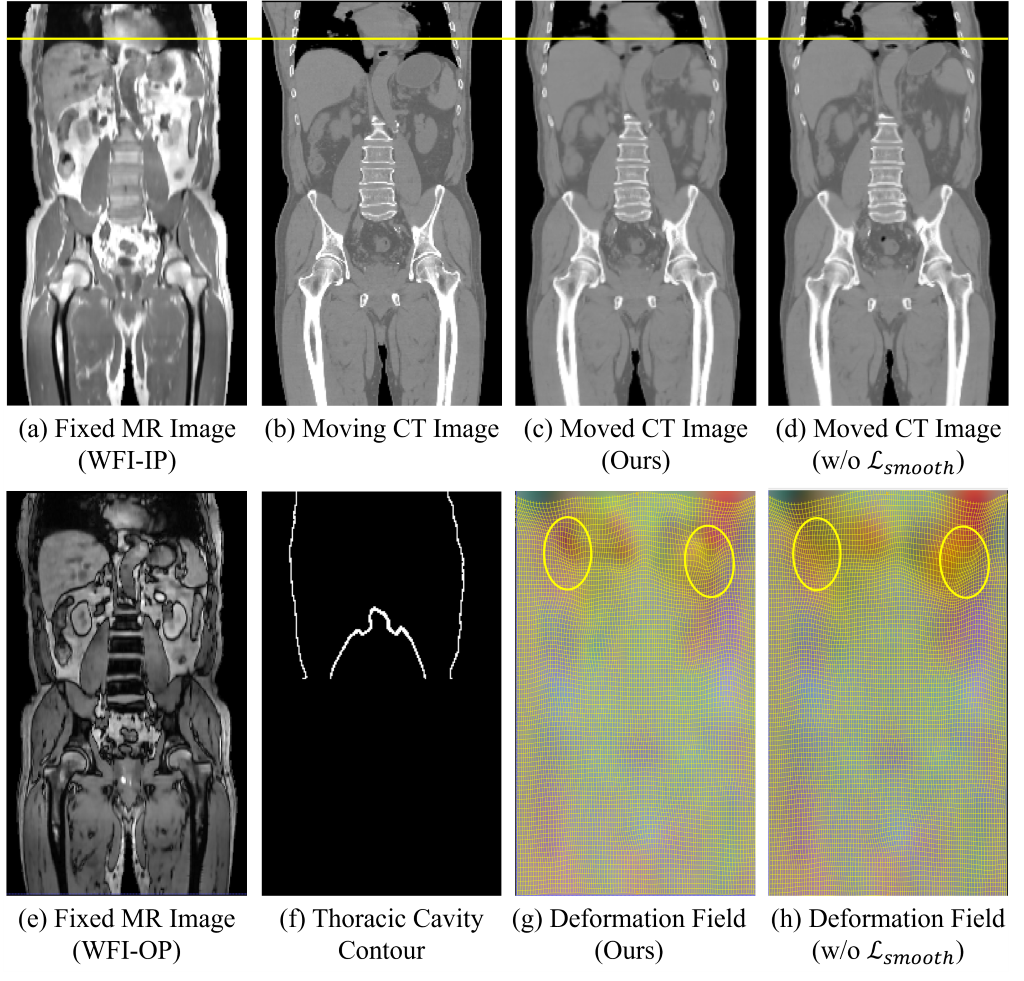}
    \caption{The respiration-aware smoothness regularization $\mathcal{L}_{smooth}$ enforces a deformation field that is less smooth near the thoracic cavity contour, thus preventing the ribs from sliding. 
    It can be observed along the yellow line that although abdominal organs are precisely registered, there is a noticeable slippage of the ribs when $\mathcal{L}_{smooth}$ is not used.}
    \label{fig:smooth}
\end{figure}

\subsubsection{Ablation Study}
To validate the effect of key components in each proposed module, we conduct an ablation study on these model configurations, including (1) the gated U-Net design in SGSyn, (2) the tissue-aware image similarity constraint $\mathcal{L}_{sim}$ and the respiration-aware smoothness regularization $\mathcal{L}_{smooth}$ in SpatAlign, and (3) the contrastive loss $\mathcal{L}_{contra}$ in SemAlign.  
We first establish a baseline framework by removing all key components.
The baseline framework consists of a synthesis model and a registration model that employ vanilla 3D U-Net architectures, and it follows our training scheme except that it adopts conventional registration loss functions.
Subsequently, we incrementally add each component to the baseline to demonstrate their effectiveness.

As shown in Fig.~\ref{fig:ablation}, the baseline result (the fourth column) exhibits a notable failure in the synthesis of bone regions, especially the ribs. 
The shapes of the generated organs are also misaligned with the input MR images.
Employing the gated U-Net as the synthesis model (the fifth column) removes many artifacts, but still falls short in generating high-quality ribs.
After adopting the proposed losses from SpatAlign, the synthetic results (the sixth column) have yielded well-aligned organ structures and successfully generated ribs.
However, the boundaries of abdominal soft tissues remain unclear.
By further integrating SemAlign that enforces organ-related semantic authenticity, our method produces the best outcomes (the last column).
The quantitative results in Table~\ref{tab:ablation} also show increased performance when adding each component to the baseline, which can further validate their contributions.

In addition to the experiments described above, we particularly visualize the effect of the respiration-aware smoothness regularization $\mathcal{L}_{smooth}$ in Fig.~\ref{fig:smooth}.
It can be observed along the yellow horizontal line that the synthetic CT livers of both results are spatially aligned with the input MR images. 
However, there is a pronounced slippage of the ribs when the points of the thoracic cavity contour are involved in smoothness regularization. 
The utilization of respiration-aware smoothness regularization $\mathcal{L}_{smooth}$ enforces a deformation field that is less smooth near the thoracic cavity contour by excluding this region from the loss calculation, thus producing physically realistic deformations for discontinuous motion.

\subsection{Evaluation of PET Attenuation Correction}
To validate the utility of our synthetic CT images for PET attenuation correction, we conducted a quantitative comparison against two alternative methods in MRAC: (1) traditional segmentation-based MRAC method using ultrashort echo time (UTE) pulse sequences~\cite{keereman2010mri}, which segments MR images into four tissue types (\ie, tissue, fat, lung, and air) and assigns corresponding attenuation coefficients; (2) sc-CycleGAN~\cite{yang2020unsupervised} which synthesizes CT images from MR images to generate AC maps.
We perform PET attenuation correction using the paired CT images to produce the ground-truth PET images for the testing data, and then calculate SUV differences between the reconstructed and ground-truth PET images across various regions of interest (ROIs).

\begin{figure}[t]
    \centering
    \includegraphics[width=\linewidth]{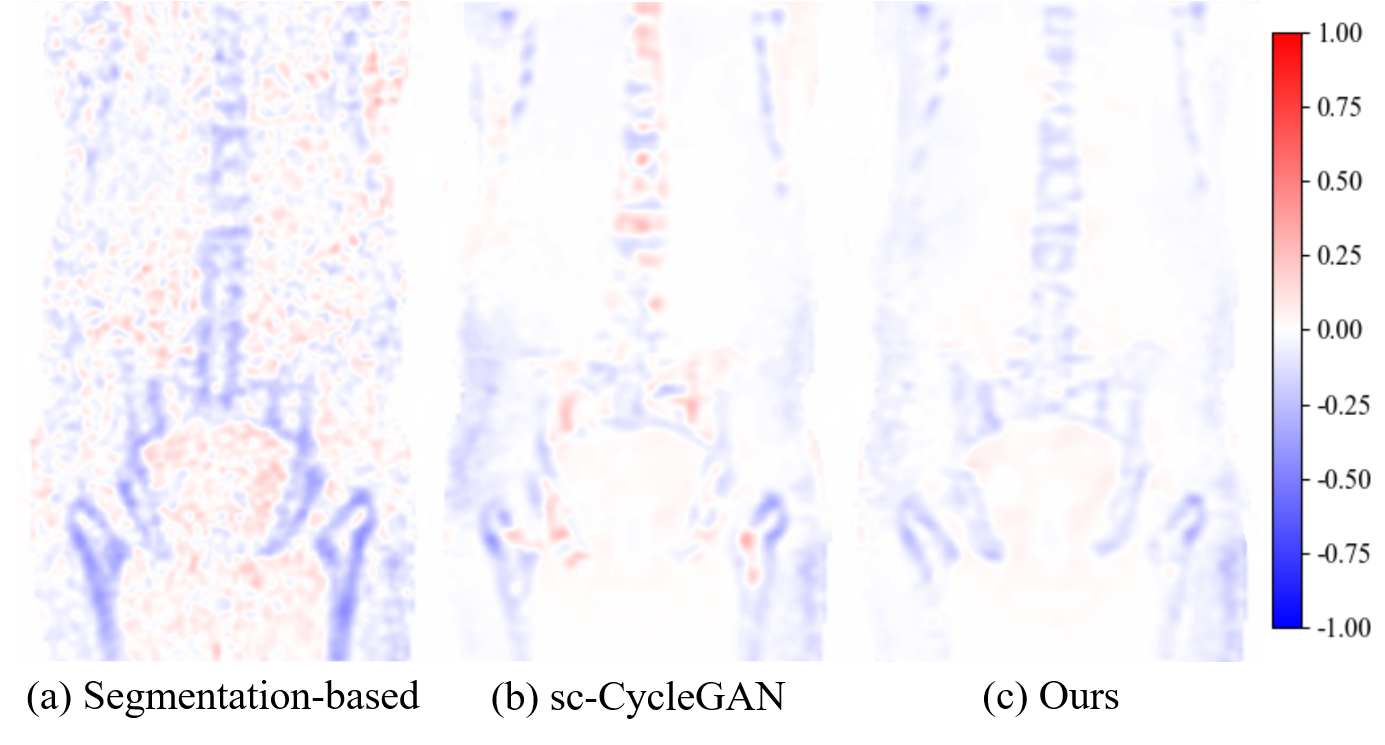}
    \caption{The SUV difference maps of reconstructed PET images using different approaches are visualized to demonstrate our method's superiority in PET attenuation correction during PET/MR imaging.}
    \label{fig:pet_recon}
\end{figure}

The results in Table~\ref{tab:pet_recon} show that the segmentation-based method yields higher SUV differences than the deep learning-based methods across all ROIs, indicating that it cannot obtain accurate tissue densities for predicting AC maps.
The PET images reconstructed using sc-CycleGAN exhibited larger SUV differences in the femurs due to its erroneous generation of bones, which is consistent with the findings of previous experiments.
Our method is more robust than sc-CycleGAN because both methods achieve similar average SUV differences but our method yields significantly lower standard deviations of SUV differences.
We also visualize their SUV difference maps of reconstructed PET images in Fig.~\ref{fig:pet_recon}.
The segmentation-based method leads to a pronounced underestimation of SUV values across all bones and produces many noises in soft tissues.
While the use of sc-CycleGAN can eliminate most noises within soft tissues, it still overestimates SUV values in the spine and the femurs.
In contrast, the reconstructed PET image using our method shows fewer errors in estimating SUV values. 
It accurately reflects the electron densities of various tissues, making it a more reliable alternative to PET attenuation correction in PET/MR imaging.

\section{Discussion and Conclusion}
Reliable PET attenuation correction is crucial for accurate PET reconstruction and quantification analysis in PET/MR imaging.
MR-to-CT synthesis emerges as a vital alternative for PET attenuation correction by predicting AC maps from synthetic CT images.
Although many MR-to-CT synthesis methods have been developed, they do not adequately address the challenges of whole-body MR-to-CT synthesis including the issue of spatial misalignment and the complexity of intensity mapping.
In this paper, we propose a novel whole-body MR-to-CT synthesis framework consisting of three modules that collectively address these challenges, and have conducted extensive experiments to showcase its great potential to advance PET/MR imaging techniques.

The primary challenge for whole-body MR-to-CT synthesis is the spatial misalignment during the acquisition of paired MR and CT images.
Previous studies mainly use CycleGAN-based methods to circumvent the need for image registration.  
However, these methods are designed for body sub-regions where the spatial misalignment issue is less pronounced.
Besides, such a cycle-loop framework solely relies on adversarial training to learn CT intensity distribution, which is prone to generating implausible anatomical structures.
In contrast, our proposed SpatAlign meticulously accounts for the impacts of tissue variety and respiration movements, emerging as a proactive solution to the problem of spatial misalignment.
The experimental results demonstrated that adopting two well-designed loss functions in SpatAlign can greatly improve the registration accuracy of anatomical structures, thus enhancing the reliability of whole-body MR-to-CT synthesis.

Another challenge is the complex intensity mapping relationship due to the inherent difficulties in aligning the distinct imaging characteristics of the two modalities.
The substantial disparity in the intensity distribution between MR and CT images leads to undesired artifacts during the synthesis process. 
The proposed SGSyn employs structure-guided attention gates to constrain the boundaries of anatomical structures in the synthetic CT images, effectively reducing many potential artifacts as evidenced by our experimental results. 
However, it encounters limitations when attempting to process soft tissues in the abdominal region, where edges are not readily discernible in CT images. 
In this way, SemAlign is proposed to exploit semantic information to complement SGSyn through contrastive learning, thereby enhancing the overall image quality.
Our experiments illustrated that the synergy of SGSyn and SemAlign can adeptly tackle the intensity mapping issue in whole-body MR-to-CT synthesis.


Our current study still has some limitations that will be addressed in future work.
First, our method necessitates the use of paired data during training, which imposes strict requirements on data collection and consequently limits the scope of our application.
We intend to leverage some adaptive approaches to reduce our reliance on paired data in the future.
Second, SemAlign demands high-precision segmentation to extract organ-related features. 
However, occasional poor segmentation outcomes can compromise the efficacy of contrastive learning.
To mitigate this issue, we plan to implement contrastive learning in an unsupervised manner to alleviate the dependency on segmentation in future work, aimed at enhancing the robustness of semantic alignment.

\bibliographystyle{IEEEtran}
\bibliography{ref.bib}
\end{document}